\documentclass[letterpapaer,journal,twoside]{IEEEtran}
\usepackage[dvipdfmx]{graphicx}
\usepackage{amsmath}
\usepackage{amsfonts}
\usepackage{amssymb}
\usepackage{bm}

 \def\hugesymbol#1{\mbox{\strut\rlap{\smash{\Huge$#1$}}\quad}}
\newcommand{\bigzerol}{\smash{\lower1.0ex\hbox{\bg 0}}}

\begin{document}
\title{\fontsize{16pt}{0pt}\selectfont Overlap Frequency Domain Equalization for Faster-than-Nyquist Signaling}
\author{Hiroyuki~Fukumoto and~Kazunori~Hayashi,~\IEEEmembership{Member,~IEEE}
\thanks{The authors are with the Graduate School of Informatics, Kyoto University, Kyoto, 606-8501 Japan e-mail: 
(fukumoto.h@sys.i.kyoto-u.ac.jp; kazunori@i.kyoto-u.ac.jp).}
}

%
%

\markboth{IEEE communications letters, VOL.~x, NO.~y, MONTH~2015}
{fukumoto and hayashi: Overlap frequency domain equalization for faster-than-Nyquist signaling}
%

\IEEEpubid{0000--0000/00\$00.00~\copyright~2012 IEEE}


\maketitle

\begin{abstract}
This letter proposes the Faster-than-Nyquist signaling (FTNS) using overlap frequency domain equalization (FDE),
which compensates the inter-symbol interference (ISI) due to band limiting filters of the FTNS at the transmitter and
the receiver as well as the frequency selective fading channel.
Since overlap FDE does not require any guard interval (GI) at the transmitter such as cyclic prefix (CP), 
higher spectral efficiency can be achieved compared to FTNS scheme using the conventional FDE.
In the proposed method, the equalizer weight is derived based on minimum mean square error (MMSE) criterion 
taking the colored noise due to the receiving filter into consideration.
Moreover, we also give an approximated FDE weight in order to reduce the computational complexity.
The performance of the proposed scheme is demonstrated via computer simulations.
\end{abstract}

\section{Introduction}
\IEEEPARstart{F}{aster}-than-Nyquist signaling (FTNS)~\cite{FTNO}~\cite{FTNU} is considered as one of key technologies for the fifth generation (5G) mobile communications systems~\cite{5G}.
FTNS can increase the transmission rate without increasing the required frequency bandwidth by transmitted modulated symbols with the rate faster than the Nyquist rate, which ensures inter-symbol interference (ISI)-free transmission in band limiting channels. 
It has been analytically shown that the minimum Euclidean distance between sequences of binary phase shift keying (BPSK) symbols 
using sinc-pulse does not degrade even with the rate 25\% faster than Nyquist rate.
This means that, if the maximum likelihood sequence estimation (MLSE) is employed at the receiver, 
transmission rate can be increased by 25\% without the degradation of the bit error rate (BER) performance.
However, the MLSE requires high computational complexity, especially, when the ISIs due to the transmitting/receiving filters as well as the frequency selective fading channel are simultaneously compensated. \par
To reduce the complexity, the FTNS receiver using frequency domain equalization (FDE)~\cite{FDE} has been recently 
proposed as a sub-optimal method in~\cite{FDEFTN}.
Moreover, the idea of~\cite{FDEFTN} has been extended to the method using frequency domain decision-feedback equalizer in~\cite{FSEDFE}.
Furthermore, the FDE-aided soft decision and the multistage serially concatenated turbo architecture are applied for the
FTNS demodulator in~\cite{FDEAID}.
However, these conventional FDE-based receivers require the insertion of the guard interval~(GI) such as cyclic prefix (CP) 
at the transmitter, which results in the reduction of the effective transmission rate.
Moreover, the colored noise, which is inherent in the received signal of the FTNS, is not explicitly considered in~\cite{FDEFTN} 
and~\cite{FDEAID}, while ~\cite{FSEDFE} takes the colored noise into consideration but with the limitation that
the sampling rate at the receiving filter is assumed to be twice the symbol rate in the formulation.\par
\IEEEpubidadjcol
This letter proposes an FTNS transceiver using overlap FDE~\cite{OFDE1}\cite{OFDE2}, which has been originally 
proposed for the FDE systems with insufficient CP or without CP.
Since the overlap FDE does not require the insertion of any CP at the transmitter at the cost of slight increase 
in computational complexity at the receiver, the proposed scheme can improve the spectral efficiency with the moderate complexity.
Also, we adopt the FDE weight based on minimum mean square error~(MMSE) criterion in order to take into account for the impact of the colored noise, 
and derive approximated one-tap FDE weight to reduce the computational complexity for the equalization.
Computer simulation results show that there is a significant difference in BER performance between the receivers using FDE weights 
with and without considerations of the impact of the colored noise.
Moreover, we show that the proposed method can achieve better BER performance than that of the FTNS transceiver using conventional FDE 
with CP~\cite{FDEFTN} for a fixed transmission rate.
In addition, we demonstrate that the proposed scheme can achieve 30\% higher transmission rate than that of Nyquist rate without 
performance degradation, when root raised cosine (RC) filter with the roll off factor of 0.5 is used for transmitting and receiving filters.

\section{System Model}
\begin{figure}[t]
\centering
\includegraphics[width=80mm]{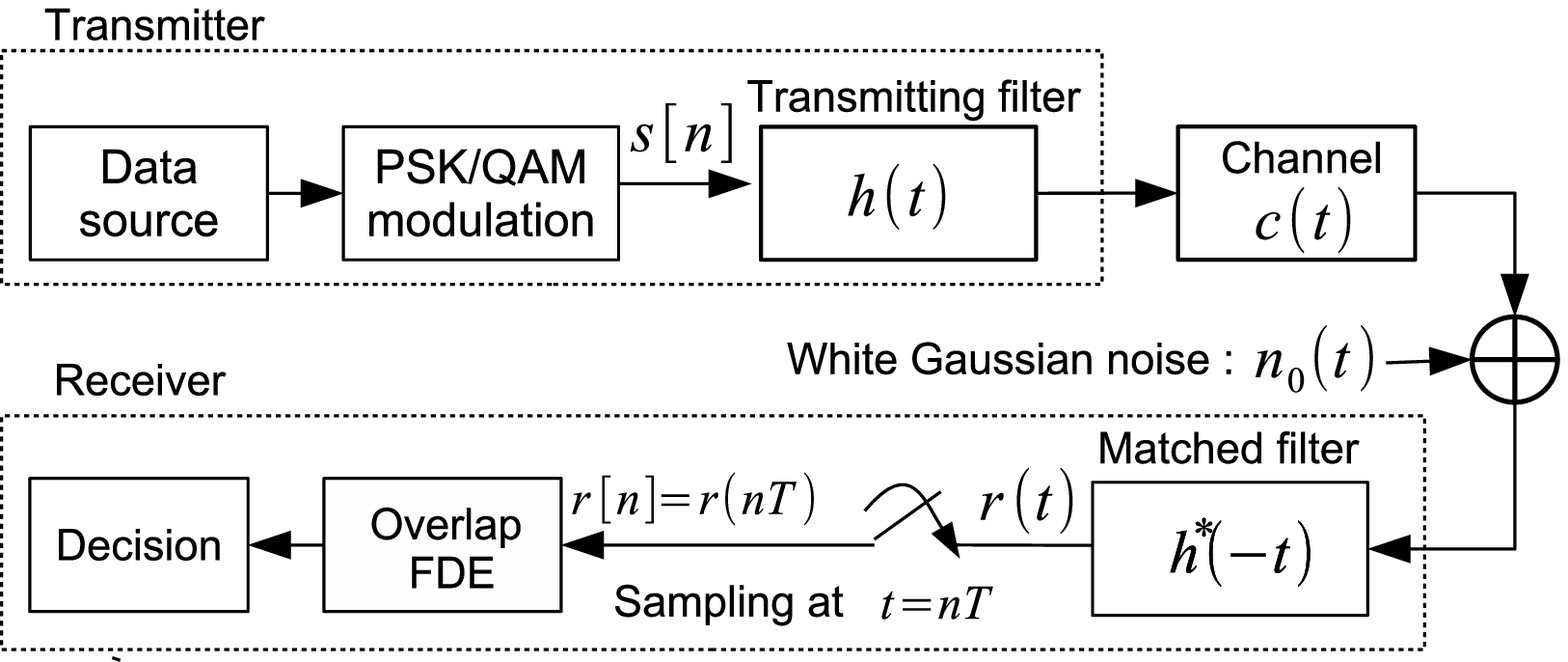}
\caption{Proposed FTNS transmission model}
\label{FTNM}
\end{figure}

Fig.\ref{FTNM} shows the configuration of the FTNS transmission system model considered in this letter.
The transmitted symbol $s[n],\ (n \in \mathbb{Z})$ is generated by baseband modulation with PSK (Phase Shift Keying) 
or QAM (Quadrature Amplitude Modulation) using the binary sequence from the data source. 
We assume $E\{s[n]\}=0$ and $E\{s[n]s^*[m]\}=\sigma_{\mathrm{s}}^2\delta[n-m]$, where $E\{\cdot\}$ represents the expectation, 
superscript $*$ a complex conjugate, and $\delta[n]$  Kronecker delta. 
The sequence of $s[n]$ is transmitted after passing through the transmitting filter $h(t)$ at the interval of $T$.
Here, we assume $h(t)$ to be the root Nyquist filter with $\int_{-\infty}^{\infty}|h(t)|^2 dt =1$, and
$g(t)=\int_{-\infty}^{+\infty}h(\tau)h^{*}(\tau-t)~d\tau$ satisfies the Nyquist criterion for the symbol period of of $T_0$.
In FTNS, in order to increase the transmission rate, $T$ is set to be less than $T_0$. Thus, we have $T=\gamma T_0,\ (0 < \gamma <1)$.

The received signal after matched-filtering with $h^*(-t)$ is given by
\begin{equation}\label{recFTN}
r(t) = \sum_{m=-\infty}^{\infty}s[m] q(t-mT) + \eta (t),
\end{equation}
where $q(t)$ is the impulse response of channel including the impact of the frequency selective channel 
$c(t)(\neq 0 ~\mbox{for}~ T_{\rm min} \leq t \leq T_{\rm max})$ and
transmitting/receiving filters and is given by
\begin{equation}
q(t)= \int_{-\infty}^{\infty}g(\tau) c(t-\tau) d \tau,
\end{equation}
and $\eta(t)$ is the noise component given by 
\begin{equation}
\eta(t)= \int_{-\infty}^{\infty} n_0(\tau) h^{*}(\tau-t) d\tau,
\end{equation}
where the white noise $n_0(t)$ follows the complex-valued Gaussian distribution $\mathcal{CN}(0,N_0)$.
%
%
The received signal is sampled with the sampling period of $T$, which is the same as the symbol period, then the $n$th sample is given by 
\begin{equation}\label{FTNS1}
r[n] = r(nT)=\sum_{m=-\infty}^{\infty}q[n-m] s[m]  +\eta[n],
\end{equation}
where $q[n]=q(nT)$ and $\eta[n]=\eta(nT)$.
Note that the mean and the correlation of the sampled noise $\eta[n]$ are given by $E\{\eta[n]\} = 0$ and $E\{\eta[n]\eta^*[m]\}=N_0 g((n-m)T)$, respectively.\par

\section{Proposed FTNS Receiver with Overlap FDE}
In the proposed method, overlap FDE~\cite{OFDE1},\cite{OFDE2},\cite{OFDE3} is employed to compensate ISIs caused by band limiting filters (i.e., transmitting and receiving filters) as well as the frequency selective channel. 
Since overlap FDE performs equalization with block-wise signal processing at the receiver, while the transmitted signal need not be block-wise,
we firstly give a block representation of the received signal.
Assuming that $g(t)= 0$ for $|t|>\nu T,~\nu \in \mathbb{Z}$ as in~\cite{FDEFTN} by truncating tails of the impulse response,
a received signal block of size $N$ defined as $\bm{r}[k] := (r[k] \cdots r[k+N-1])^{\rm T}$, where the superscript T represents the transpose,
is given by
\begin{equation}\label{FTNS3}
\bm{r}[k] = \bm{Q}_0 \bm{s}[k] + \bm{Q}_1 \bm{s}[k+N]+ \bm{\eta}[k],
\end{equation}
where
\begin{equation}
\bm{Q}_0 = 
\begin{pmatrix}
q[n_{\rm max}]    & \hdots    & q[n_{\rm min}]  &                   &                \\
                        & \ddots     &                      & \ddots        &\hugesymbol{O}\\
                        &               & \ddots            &                  & q[n_{\rm min}]     \\          
                        &          &                      & \ddots        & \vdots     \\
 \hugesymbol{O}                           &               &                      &                  & q[n_{\rm max}]    \\
\end{pmatrix}
\in \mathbb{C}^{N \times N},
\end{equation}
and
\begin{equation}
\bm{Q}_1 = \left(
\begin{array}{ccccc}
                        &           &                           &         \\
                         &           &  \hugesymbol{O}                          &         \\
q_{n_{\rm min}}      &           &    &           \\
\vdots                 & \ddots &                          &            \\
q_{n_{\rm max}-1} & \hdots  & q_{n_{\rm min}}     &           
\end{array}
\right)
\in \mathbb{C}^{N \times N}.
\end{equation}
Also, $n_{\rm min}=-\nu + \lceil T_{\rm min}/T \rceil$ and $n_{\rm max}= \nu + \lfloor T_{\rm max}/T \rfloor$, where 
$\lceil x \rceil$ and $\lfloor x \rfloor$ are the smallest integer not less than $x$ and the largest integer not greater than $x$, respectively.
In addition, $\bm{s}[k] := (s[k-n_{\rm max}] \cdots s[k-n_{\rm max}+N-1] )^{\rm T}$ and $\bm{\eta}[k] := (\eta[k] \cdots \eta[k+N-1])^{\rm T}$.

By defining $\bm{Q}:=\bm{Q}_0+\bm{Q}_1$, \eqref{FTNS3} can be rewritten as
\begin{equation}\label{FTNS4}
\bm{r}[k]  = \bm{Q}\bm{s}[k] + \bm{Q}_1 (\bm{s}[k+N]-\bm{s}[k])+\bm{\eta}[k].
\end{equation}
It should be noted here that $\bm{Q}$ is a circulant matrix, which can be diagonalized by using $N$-point 
DFT (discrete Fourier transform) matrix $\bm{D}$, whose $(n,m)$ $(n,m=1,\ldots,N)$ element is $e^{-j2\pi(n-1)(m-1)/N}/\sqrt{N}$, 
as $\bm{Q}=\bm{D}^{\rm H}\bm{\Lambda}\bm{D}$, where the superscript H denotes Hermitian transpose and
$\bm{\Lambda} \in \mathbb{C}^{N \times N}$ is a diagonal matrix whose diagonal elements are the DFT of the first column of $\bm{Q}$.
Thus, if we apply FDE for $\bm{Q}$ ignoring the second term of right hand side of \eqref{FTNS4}, which represents 
inter-block interference and ISI components due to the lack of CP,
the FDE output is given by
\begin{align}\label{FTNS5}
&\bm{D}^{\rm H}\bm{W}\bm{D}\bm{r}[k] = \bm{D}^{\rm H}\bm{W}\bm{\Lambda}\bm{D}\bm{s}[k] \nonumber \\
&+ \bm{D}^{\rm H}\bm{W}\bm{D}\bm{Q}_1 (\bm{s}[k+N]-\bm{s}[k])+\bm{D}^{\rm H}\bm{W}\bm{D}\bm{\eta}[k],
\end{align}
where $\bm{W}\in \mathbb{C}^{N\times N}$ is an FDE weight matrix.

Since it is empirically known that head and tail of the signal block at the FDE output suffer from
the residual interference due to the second term of \eqref{FTNS5},
only the middle part of $\bm{D}^{\rm H}\bm{W}\bm{D}\bm{r}[k]$ is used as the output of overlap FDE.
Specifically, discarding $p$ elements from the head and $q$ elements from the tail,
the output of the overlap FDE is given by
\begin{equation}\label{OFDEeq3}
\tilde{\bm{r}}[k] = (\bm{0}_{M\times p},\bm{I}_{M},\bm{0}_{M \times q})\bm{D}^{\rm H}\bm{W}\bm{D}\bm{r}[k],
\end{equation}
where $M=N-p-q$ is the size of the overlap FDE output, $\bm{I}_{M}$ represents the identity matrix of size $M \times M$ 
and $\bm{0}_{M\times p}$ a zero matrix of size $M \times p$. 
The next block processing is performed for the received signal block of $\bm{r}[k+M] = (r[k+M] \cdots r[k+M+N-1])^{\rm T}$
in the same way as described above.

Then, we consider the FDE weight based on MMSE criterion. 
If we ignore the second interference term in \eqref{FTNS4}, the MMSE-FDE weight can be obtained by solving the following optimization problem:
\begin{align}\label{MMSES}
\bm{W}_{\rm c} = {\rm arg}\min_{\bm{W}\in \mathbb{C}^{N\times N}}E\bigl\{||\bm{D}^{\rm H}\bm{W}\bm{D}(\bm{Q} \bm{s}[k]+\bm{\eta}[k]) -\bm{s}[k]||^2 \bigr\}.
\end{align}
The MMSE-FDE weight is obtained as
\begin{equation}\label{CMMSE}
\bm{W}_{\rm c}=\bm{\Lambda}^{\rm H}\left(\bm{\Lambda}\bm{\Lambda}^{\rm H}+\frac{1}{\sigma_{\mathrm{s}}^2}\bm{D}E\{\bm{\eta}[k]\bm{\eta}^{\rm H}[k]\}\bm{D}^{\rm H} \right)^{-1}.
\end{equation}
Note that since $\bm{D}E\{\bm{\eta}\bm{\eta}[k]^{\rm H}[k]\}\bm{D}^{\rm H}$ is not diagonal due to the colored noise, 
$\bm{W}_{\rm c}$ is also non-diagonal.
Thus, MMSE-FDE is not achieved with one-tap FDE, which may spoil the motivation to utilize FDE because of the high computational complexity.

In order to reduce the complexity, the proposed FTNS scheme obtains the FDE weights by approximating $\bm{D}E\{\bm{\eta}\bm{\eta}[k]^{\rm H}[k]\}\bm{D}^{\rm H}$  with a diagonal matrix, whose diagonal entries are the same as those of $\bm{D}E\{\bm{\eta}\bm{\eta}^{\rm H}]\bm{D}^{\rm H}$.
Then, the approximated FDE weight of $\bm{W}_{\rm c}$ can be given by
\begin{equation}\label{CDMMSE}
\tilde{\bm{W}}_{\rm c} = \bm{\Lambda}^{\rm H}\left( \bm{\Lambda}\bm{\Lambda}^{\rm H}+\frac{1}{\sigma_{\mathrm{s}}^2}\bm{\Phi}_{\eta} \right)^{-1},
\end{equation}
where $\bm{\Phi}_{\eta}={\rm diag}(\Phi_{\eta}[0],...,\Phi_{\eta}[N-1])$ is a diagonal matrix composed by the power spectrum density of the noise
\begin{eqnarray}
\Phi_{\eta}[n] 
&=&\frac{N_0}{N}\sum_{l=0}^{N-1}\sum_{m=0}^{N-1}g((l-m)T)e^{-j\frac{2\pi (l-m)n}{N}}. \label{CFCW}
\end{eqnarray}

Note that, if we solve \eqref{MMSES} with the assumption of white noise, we obtain FDE weight of
\begin{equation}
\bm{W}_{\rm w} = \bm{\Lambda}^{\rm H}\left( \bm{\Lambda}\bm{\Lambda}^{\rm H}+\frac{N_0}{\sigma_{\mathrm{s}}^2}\bm{I}_{N} \right)^{-1},
\end{equation}
which is widely employed in conventional FDE systems including ~\cite{FDEFTN}.

\section{Simulation results}
\begin{figure}[tbp]
\begin{center}
\includegraphics[width=85mm]{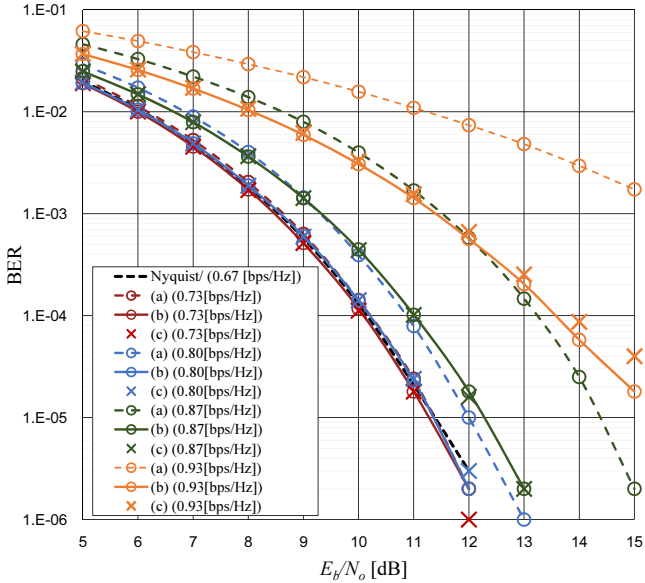}
\caption{BER performance of FTNS using FDE with CP for different weights}
\label{result10}
\end{center}
\end{figure}
We evaluate the BER performance via computer simulations in order to confirm the effectiveness of the proposed method.
In all the simulations, binary PSK is employed for baseband modulation, and root RC filter with roll off factor 0.5 is
used for transmitting and receiving filters. Thus, the Nyquist rate corresponds to $0.67$ [bps/Hz].
For FTNS, transmission rates are set to be $10\%$ (0.73 [bps/Hz]), $20\%$ (0.80 [bps/Hz]), $30\%$ (0.87 [bps/Hz]) and $40\%$ (0.93 [bps/Hz]) 
higher than the Nyquist rate.
We evaluate the averaged BER by counting error using symbol by symbol detection after compensating ISI.
The DFT size of FDE is set to be 512 and, in the block processing at the receiver, $g(t)$ is truncated with $\nu=10$.\par

In order to confirm the impact of the colored noise, we firstly compare the BER performance of FTNS transmissions in AWGN
 (Additive White Gaussian Noise) channel using FDE with CP of length 20 symbols
for different FDE weights in Sect.III, i.e.,  $({\rm a})\bm{W}_{\rm w},\ ({\rm b})\bm{W}_{\rm c}$, and $({\rm c})\tilde{\bm{W}}_{\rm c}$. 
Fig. \ref{result10} shows the BER performance against $E_b/N_0$.
Here, $\gamma$ is set to be $\gamma=\{0.875,~0.802,~0.740,~0.687\}$, which respectively correspond to rates of 
$\{0.73,~0.80,~0.87,~0.93\}$ [bps/Hz], where the rate loss due to CP is taken into consideration.
From Fig. \ref{result10}, we observe the BER performance of (b) is significantly better than that of (a) for the same transmission rate
especially for smaller $\gamma$, which implies the importance to consider the impact of the colored noise.
Also, we can observe that (c) can achieve comparable performance to (b) for the same transmission rate,
which demonstrates the validity of the approximation in the derivation of (c).

\begin{figure}[tbp]
\begin{center}
\includegraphics[width=85mm]{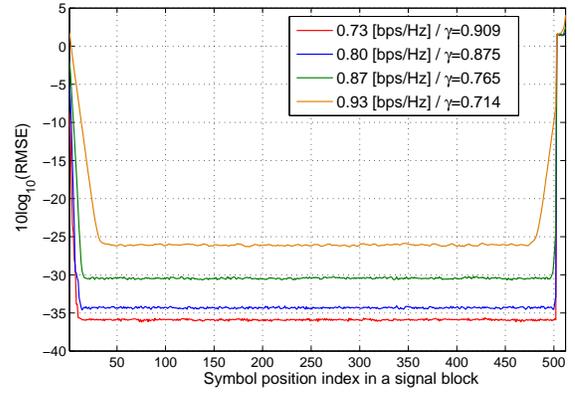}
\caption{RMSE between FDE output and transmitted symbol for each symbol position in a block}
\label{result20}
\end{center}
\end{figure}

Then, we demonstrate the validity of the proposed approach using overlap FDE for the FTNS transmission.
Fig. \ref{result20} shows the root MSE (RMSE) between the FDE output using the weight (c) and the actual transmitted symbol averaged over 1,000 FDE outputs
versus the symbol position in a signal block, where the additive noise power $N_0$ is set to be 0 and the channel is flat with a fixed gain of 1.
Here, $\gamma$ is set to be $\{0.909,~0.833,~0.765,~0.714\}$, which correspond to rates of $\{0.73,~0.80,~0.87,~0.93\}$ [bps/Hz], respectively.
From the results, we can see that RMSE values at the middle part of the block are significantly smaller than the head and the tail of the block for all the values of $\gamma$, 
and that we only have to remove rather small portions of the FDE output, which means the increase in computational complexity at the receiver due to
the employment of overlap FDE is rather limited.\par

\begin{figure}[tbp]
\begin{center}
\includegraphics[width=85mm]{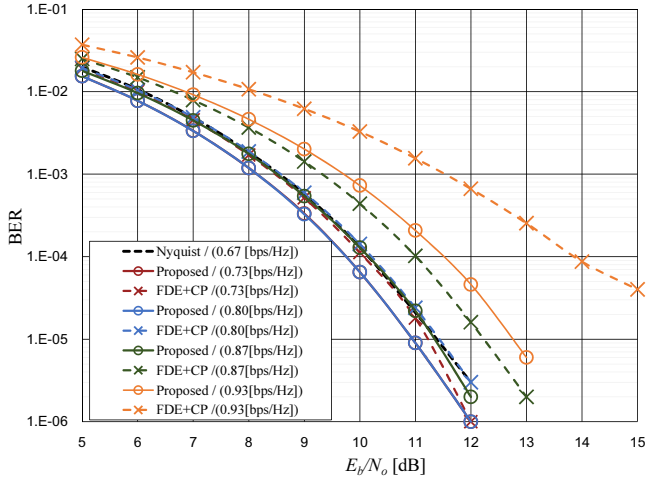}
\caption{BER performance comparison between the proposed FTNS scheme and FDE with CP~\cite{FDEFTN} ((I): AWGN channel)}
\label{result21}
\end{center}
\end{figure}

\begin{figure}[tbp]
\begin{center}
\includegraphics[width=85mm]{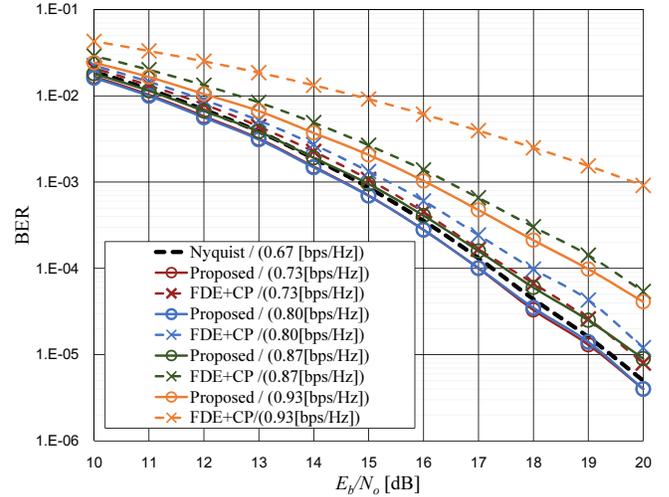}
\caption{BER performance comparison between the proposed FTNS scheme and FDE with CP~\cite{FDEFTN} ((I\hspace{-,1em}I): 10tap Rayleigh fading channel)}
\label{result22}
\end{center}
\end{figure}

Figs. \ref{result21} and \ref{result22} show the BER performance of the proposed FTNS scheme using overlap FDE in AWGN channel and 
10-tap frequency selective Rayleigh fading channel having maximum duration of $16T$, respectively, for different values of $\gamma$ used in Fig. \ref{result20}.
Also, the performance of FTNS using FDE with CP~\cite{FDEFTN} is also plotted in the same figures for a comparison purpose.
In both schemes, the FDE weight of (c) is employed to cope with the colored noise, while (a) is used in the original scheme in~\cite{FDEFTN}.
In the proposed FTNS scheme, $p$ and $q$ are set to be $p=q=30$ for the AWGN channel and $p=q=128$ for the frequency selective fading channel.
In the scheme using FDE with CP, the length of CP is set to be 20 for the AWGN channel and 36 for the frequency selective fading channel,
and $\gamma$ is set to be $\{0.875,~0.802,~0.740,~0.687\}$ for the AWGN channel and $\{0.849,~0.779,~0.719,~0.667\}$ for the frequency selective fading channel,
which results in the same effective transmission rates as those of the proposed FTNS scheme.
From the results in Figs. \ref{result21} and \ref{result22}, we can observe that the proposed FTNS scheme can achieve better BER performance than that of the scheme in~\cite{FDEFTN}
for the same rate.
Moreover, in both figures, the BER performance of the proposed FTNS scheme with the rate less than or equal to 0.87 [bps/Hz] is comparable to that 
of the Nyquist rate transmission, which means that $30\%$ higher transmission rate than the Nyquist rate transmission
can be achieved with the proposed FTNS scheme without BER performance degradation if root RC filter with roll off factor 0.5 is employed for transmitting and receiving filters.


\section{Conclusion}
In this letter, we have proposed the ISI compensation method using overlap FDE for FTNS transmission.
In order to take account for the impact of colored noises due to receiving filter, we have proposed an approximated MMSE weight for the one-tap FDE.
Computer simulation results show that the colored noise should be explicitly considered in the FTNS using FDE.
Moreover, we have confirmed that, with the proposed scheme, 30\% higher rate than the Nyquist rate can be realized without BER degradation
root RC filter having roll off factor 0.5 is used as band limiting filters.

\section*{Acknowledgements}
This work was supported in part by JSPS KAKENHI Grant Number 15K06064.

\end{document}